\documentclass{article}
\usepackage[frozencache]{minted}
\usepackage{tenor2025}
\usepackage{ifpdf}
\usepackage[english]{babel}
\usepackage{balance}

\usepackage{amsmath}
\usepackage{amssymb}
\usepackage[font=footnotesize,caption=false]{subfig}

\def\papertitle{Detecting notational errors in digital music scores}
\def\firstauthor{Léo Géré}
\def\secondauthor{Nicolas Audebert}
\def\thirdauthor{Florent Jacquemard}

\newif\ifpdf
\ifx\pdfoutput\relax
\else
   \ifcase\pdfoutput
      \pdffalse	
   \else
      \pdftrue
\fi

\ifpdf 
  \usepackage[pdftex,
    pdftitle={\papertitle},
    pdfauthor={\firstauthor, \secondauthor, \thirdauthor},
    bookmarksnumbered, 
    pdfstartview=XYZ 
   ]{hyperref}

  \usepackage[pdftex]{graphicx}
  \graphicspath{{./figures/}}
  \DeclareGraphicsExtensions{.pdf,.jpeg,.png}

  \usepackage[figure,table]{hypcap}

\else 
  \usepackage[dvips,
    bookmarksnumbered, 
    pdfstartview=XYZ 
  ]{hyperref}  
  \usepackage[dvips]{epsfig,graphicx}
  \graphicspath{{./figures/}}
  \DeclareGraphicsExtensions{.eps}

  \usepackage[figure,table]{hypcap}
\fi

\hypersetup{
    colorlinks,%
    citecolor=black,%
    filecolor=black,%
    linkcolor=black,%
    urlcolor=black
}

\usepackage{microtype}
\usepackage[table]{xcolor}
\usepackage{pmboxdraw}
\usepackage[capitalise]{cleveref}
\usepackage[export]{adjustbox}
\usepackage[edges]{forest}
\usepackage{xfrac}
\usepackage{musicography}
\usepackage{float}
\usepackage{minted}
\usepackage{booktabs}
\usepackage{tabularx}
\newcolumntype{Y}{>{\centering\arraybackslash}X}
\usepackage{graphicx}
\usepackage{array} 
\usepackage{cellspace} 
\usepackage{multicol}
\usepackage{enumitem}
\setlist[itemize]{itemsep=0pt, topsep=1pt}
\forestset{
  default preamble={
    for tree={
      folder,
      grow'=0,
      fit=rectangle,
      s sep=0,
      inner xsep=2,
      inner ysep=0,
      align=left,
      delay={
        if={or(strequal(content(), "OR"), strequal(content(), "AND")}{
            content=\texttt{\textbf{#1}},
        }{}
      },
      if n children=0{child anchor=mid west}{},
    }
  }
}

\newcommand{\xmltag}[1]{\texttt{<#1>}}
\newcommand{\token}[1]{\textbf{\texttt{#1}}}
\newcommand{\topstack}[1]{\operatorname{top}\hspace{-1.5pt}\left(#1\right)}
\newcommand{\tabitem}{\quad\llap{\textbullet}~~}

\makeatletter
\g@addto@macro{\UrlBreaks}{\UrlOrds}
\makeatother

\usepackage{xspace}
\makeatletter
\DeclareRobustCommand\onedot{\futurelet\@let@token\@onedot}
\def\@onedot{\ifx\@let@token.\else.\null\fi\xspace}
\def\eg{\emph{e.g}\onedot} 
\def\ie{\emph{i.e}\onedot} 
 
\def\etc{\emph{etc}\onedot}

\makeatother

\title{\papertitle}

 \threeauthors
   {\firstauthor}{Cnam, Cedric, Paris, France \\ %
     {\tt \href{mailto:leo.gere@lecnam.net}{leo.gere@lecnam.net}} \\ \vphantom{ }}
   {\secondauthor}{IGN, LASTIG, Saint-Mandé, France \\ Cnam, Cedric, Paris, France \\ %
     {\tt \href{mailto:nicolas.audebert@ign.fr}{nicolas.audebert@ign.fr}}}
   {\thirdauthor}{INRIA, Paris, France \\ Cnam, Cedric, Paris, France \\ %
     {\tt \href{mailto:florent.jacquemard@inria.fr}{florent.jacquemard@inria.fr}}}

\begin{document}

\capstartfalse
\maketitle
\capstarttrue

\begin{abstract}
Music scores are used to precisely store music pieces for transmission and preservation.
To represent and manipulate these complex objects, various formats have been tailored for different use cases. While music notation follows specific rules, digital formats usually enforce them leniently.
Hence, digital music scores widely vary in quality, due to software and format specificity, conversion issues, and dubious user inputs.
Problems range from minor engraving discrepancies to major notation mistakes.
Yet, data quality is a major issue when dealing with musical information extraction and retrieval.
We present an automated approach to detect notational errors, aiming at precisely localizing defects in scores.
We identify two types of errors: \textit{i)} rhythm/time inconsistencies in the encoding of individual musical elements, and \textit{ii)} contextual errors, \ie notation mistakes that break commonly accepted musical rules.
We implement the latter using a modular state machine that can be easily extended to include rules representing the usual conventions from the common Western music notation.
Finally, we apply this error-detection method to the piano score dataset ASAP \cite{foscarin:asap-dataset}. We highlight that around 40\% of the scores contain at least one notational error, and manually fix multiple of them to enhance the dataset's quality.
\end{abstract}

\section{Introduction}

Sheet music scores are, since centuries, an essential medium for the exchange of information between composers, performers, teachers, students, scholars, \etc.
We focus in particular on common Western music notation.
Scores are rich and complex hierarchical objects, containing information about notes, rhythms, meter, dynamics, expression\dots
Various encodings have been proposed over the years to represent these musical data digitally, such as **kern \cite{kern}, ABC notation \cite{abc}, MusicXML \cite{good:musicxml,musicxml}, MEI \cite{roland:mei,mei}.

As complex as they can be, scores must follow a certain number of syntactic rules to be correctly analyzed, or for their content to be leveraged in different applications, including graphical rendering \cite{read:music-notation,gould:behind-bars,brodsky:mental-representation}.
For example, a half note has twice the duration of a quarter note; notes in a same voice cannot overlap each other unless they form a chord, \ie a group of simultaneous notes of the same duration.
Score encoding formats are usually lax regarding compliance with these rules, which can lead to unconventional scores difficult (or even impossible) to work with.
In particular, verbosity and information redundancy in these formats may be sources of inconsistencies.
For example, in MusicXML, note durations are expressed both using symbols (quarter/half/eighth\dots, eventual dots) and numerical values (number of quarter notes on a timeline). However, the specification does not enforce any consistency between those, which can lead to issues in voice alignment and parsing difficulties.
Those notational errors\footnote{Composers do sometimes voluntary break some music notation rules \eg for artistic reasons, but we will consider those as errors in this work.}
can come from music editing software export, from conversion between different encoding formats, or from human encoders themselves.
Differences in encoding have an impact on music analysis tasks \cite{napoles:encoding-matters}, supporting the importance of having correct encoding to effectively leverage musical scores.
Clean data is also primordial to train machine learning models \cite{sedir:data-quality-impact-ml}, especially models producing complex structured output such as score data, in the context of tasks such as Automatic Music Transcription or Optical Music Recognition.

\begin{figure}
    \centering
    \subfloat[MuseScore 4.5.2]{%
        \includegraphics[height=22px,valign=t]{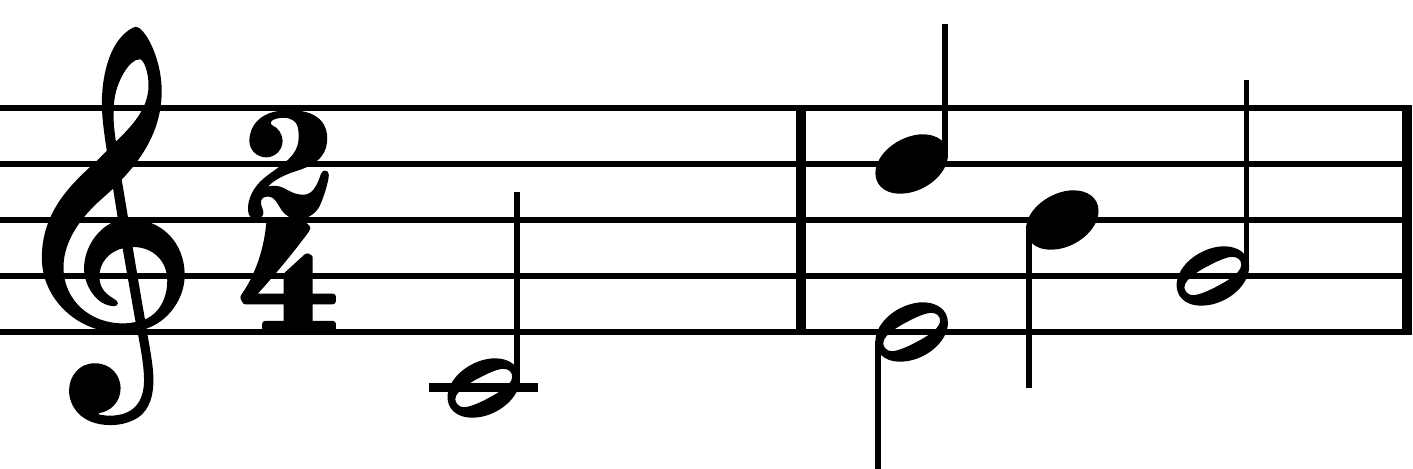}
    } \hfill
    \subfloat[Dorico 6.0.22]{%
        \includegraphics[height=22px,valign=t]{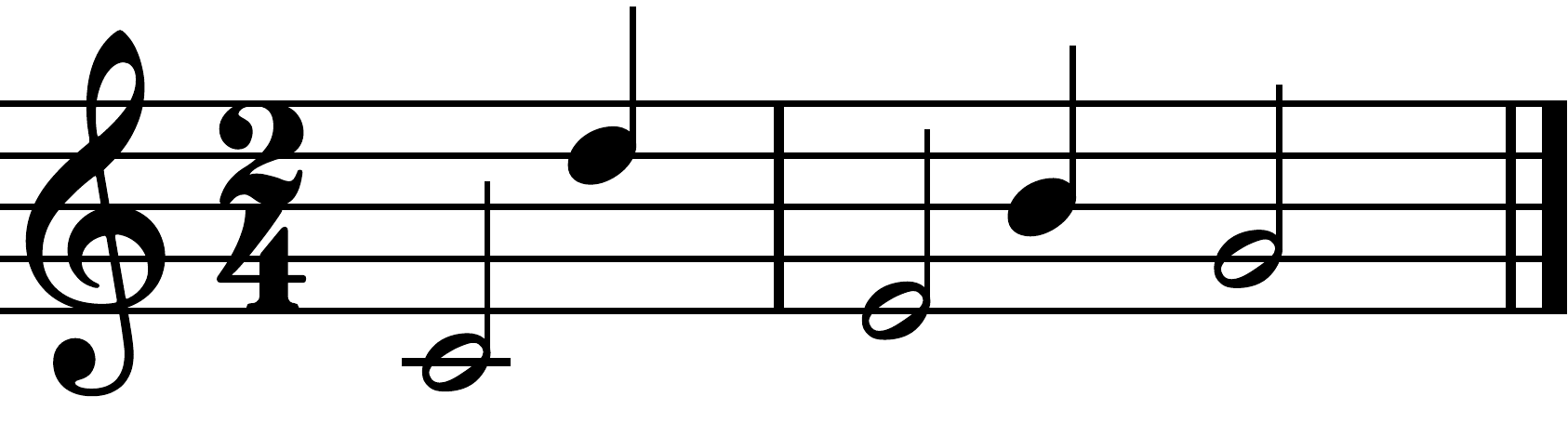}
    } \hfill
    \subfloat[Verovio 4.2.0]{%
        \includegraphics[height=22px,valign=t]{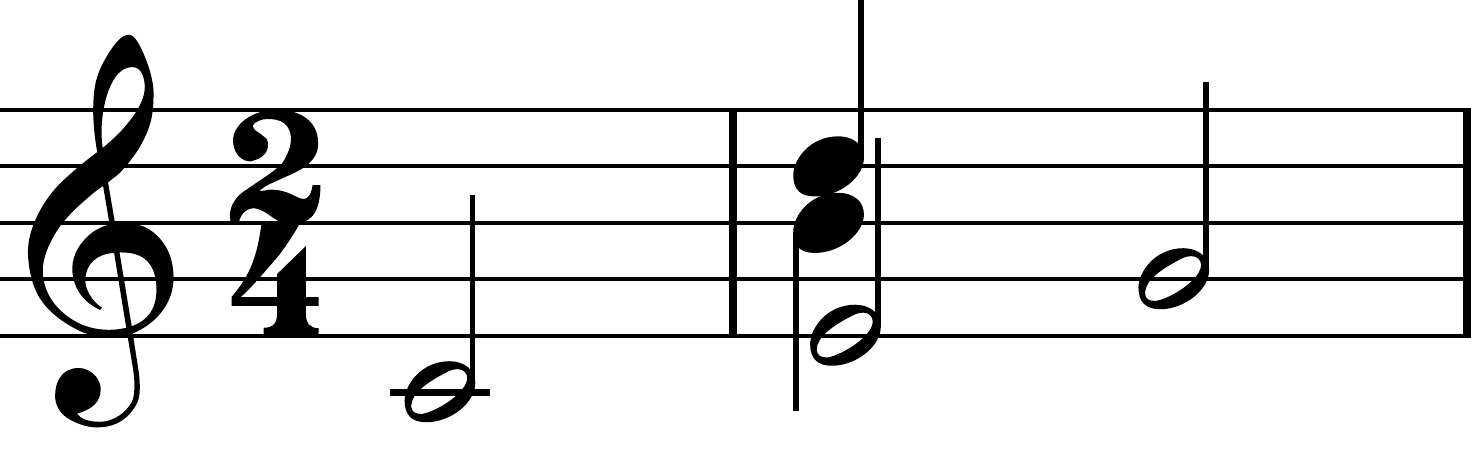}
    }
    \caption{Example of the same erroneous MusicXML file (presented in \cref{apx:error-musicxml}) rendered by three different score rendering/editing software applications.}
    \label{fig:incorrect-musicxml}
\end{figure}

All music editing software perform validation, \ie checking respect of the MusicXML specification. However, some also validate the musical content itself, looking for inconsistencies.
Furthermore, programs sometimes applies automated error-fixing mechanisms, \eg by raising explicit errors to the users, or sometimes by silently ``correcting'' the score.
Yet, those automated checks are often limited and software-specific.
As shown in \cref{fig:incorrect-musicxml}, different software handles errors differently.
In practice, we argue that it might be easier to point the error to the user, and let them fix it themselves rather than guessing their initial intention.

To evaluate the quality of digital score libraries containing scores from different sources, formats, and using different conventions, the GioQoso Project \cite{fiala:data-quality2018,foscarin:data-quality2018,foscarin:data-quality2021} proposes a quality model evaluating diverse aspects of the scores.
It investigates a wide range of potential quality issues: score content issues, engraving issues, metadata issues, \etc.
In this work, we only consider the \textit{musical content} of a score independently from any engraving and graphical rendering concerns. In other words, we only deal with the logical structure of the score, without considering the exact placement of elements relative to each other on a printed sheet.

In this article, we introduce a two-step error detection method that aims at precisely localizing notational mistakes in a score.
First, we perform a rhythm/time consistency check, that detects musical elements whose theoretical duration, as described by their symbol, is inconsistent with the actual duration specified in the MusicXML tags.
See \cref{sec:individual-errors} for the description of this kind of errors and \cref{sec:individual-errors-method} for their detection.
We chose to focus on MusicXML, as it is one of the most common open digital music format.
In the second step, we ensure that the score complies with a set of conventional music notation rules, see \cref{sec:contextual-errors,sec:contextual-errors-method}.
Our validation algorithm is flexible, built upon a set of music tokens and a state machine that could easily be extended with new rules and conventions.
This second step is format-agnostic and could be extended to any encoding, provided that a tokenizer is implemented for the required format.
We apply this method to the ASAP piano scores dataset \cite{foscarin:asap-dataset}, and identify problematic sections in approximately 40\% of the scores (\cref{sec:application}). 
Our manual qualitative analysis shows that these are notational mistakes, due to annotator error and poor software exports.
We also improve the overall quality of the ASAP dataset by fixing some of these errors.

\section{Error Types}

There are multiple sources and types of errors, and we consider the following two categories: individual errors, and contextual errors.
Individual errors refer to errors on single elements that we can detect on their own, without considering their interpretation context.
Contextual errors refer to elements that are correct on their own, but break musical rules when taking into account their musical surrounding, \eg overlapping notes in a voice.

\subsection{Individual Errors}\label{sec:individual-errors}

The errors presented in this subsection are specific to MusicXML due to how this format encodes some information.

\subsubsection{Duration Representation in MusicXML}

In MusicXML, durations are expressed as a number of quarter notes on a timeline, represented as exact fractions.
The denominator is encoded at the measure level in the integer \xmltag{divisions} tag.\footnote{Usually, only one single common denominator is used for the whole score, introduced in the first measure.}
The numerator is stored in the individual elements, in an integer \xmltag{duration} tag.
For instance, in a measure with an attribute \xmltag{divisions} of 24, a note with a tag \xmltag{duration} of 48 would last the duration of $\frac{48}{24} = 2$ quarter notes.
This fraction represents the actual duration of the note on the timeline, but is incomplete with respect to the symbology of the note \textit{value}, \ie whether it is a quarter note, a half note, \etc, and whether it has augmentation dots, or belong to a tuplet. 
For these purposes MusicXML also includes respectively the \xmltag{type}, \xmltag{dot} and \xmltag{time-modification} tags.

\subsubsection{Inconsistencies in Redundant Duration Information}

Individual errors often stem from inconsistencies in redundant information, typically in MusicXML's double encoding of the duration.
As just explained, a note with a \xmltag{duration} of 48 in a 24-division-based MusicXML, has a duration of two quarters notes, or a half note.
Yet, the MusicXML specification does not enforce that the \xmltag{type} tag represents a half note.
Therefore, some notes can have a timeline duration that is inconsistent with their theoretical duration. 
Parsers then have to decide which one to use.

These inconsistencies are generally rooted in attempts to improve the user experience, \eg by writing the score so that its engraving or MIDI playback matches the user's expectations -- at the expense of the musical content correctness. Grace notes are a typical example when they are written as normal notes. Ornaments are another one when added as invisible notes.

Tuplets are also a source of duration inconsistencies, due to rounding errors.
MusicXML theoretically allows representing any fractional duration by setting an appropriate \xmltag{divisions} tag.
Yet, some software exports scores with a default value when faced with unusual tuplets, \eg 480 for MuseScore up to version 4.4.
Because some notes durations cannot be expressed as a fraction with this denominator, their \xmltag{duration} tag is rounded to the nearest integer resulting in quantization errors.
For example, a note in a septuplet of eighth notes has a theoretical value of $\frac12 \times \frac87 = \frac47 \approx 0.5714$ quarter notes, while in a 480-division-based MusicXML file it would have a duration of $\frac{\text{round}(\sfrac47 \times 480)}{480} = \frac{274}{480} \approx 0.5708$ quarter notes.

\subsubsection{Dubious \xmltag{backup} and \xmltag{forward} Tags}

MusicXML uses \xmltag{backup} and \xmltag{forward} tags to align the different voices on the timeline.
For example, once a voice has been completed, a \xmltag{backup} tag is used to ``rewind'' the timeline back to the start of the measure and start filling a new voice.
Similarly, \xmltag{forward} tags are used to artificially ``fast-forward'' in the measure, generally in the case of incomplete voices.
Like notes, these tags have a duration indicating the displacement on the timeline.

Although there are no set rules constraining the duration of these tags, some unusual values are likely to indicate an underlying issue.
Consider a 480-division-based MusicXML file, in which a \xmltag{forward} tag has a duration of 1. This means stepping forward on the timeline by a duration of a 1024th note in a triplet, nested in a quintuplet.
This is not strictly speaking forbidden, although it is unlikely.
Music editing software typically uses such \xmltag{forward}/\xmltag{backup} tags to fill small gaps due to the accumulation of rounding errors in case of the ill-chosen \xmltag{divisions} tags discussed above.

\subsection{Contextual Errors}\label{sec:contextual-errors}

The errors presented so far were errors on their own and did not need any context to be qualified as errors.
However, some errors are more complex and do require information about surrounding notes to be properly identified.

\subsubsection{Voice Overlap Errors}

\begin{figure}
    \centering
    \includegraphics[width=0.7\linewidth]{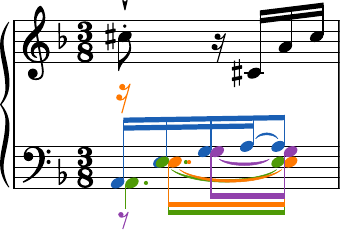}
    \vspace{-0.5em}
    \caption{Example of overlapping notes in measure 356 of 3rd movement of Beethoven's Piano Sonata No. 17 in the ASAP dataset \cite{foscarin:asap-dataset}. The notes are colored by voice. The notes in the green voice overlap with each other: no other note should start on that voice after the first dotted quarter note. This issue might come from a conversion error in the history of the score, merging two voices into one.}
    \label{fig:overlapping-notes}
\end{figure}

We consider a voice to be a succession of notes, without any gap or overlap except for chords.
In polyphonic music, a score can contain multiple voices in parallel, emphasizing different melodic lines.
Voices can sometimes be \textit{incomplete}, for example, they can start in the middle of the measure, along with an already started voice.
In this article, we only consider \textit{complete} voices, \ie without any gaps.
In case of incomplete voices, we pad them with rests.

Typical errors in voices are overlapping notes, as illustrated by \cref{fig:overlapping-notes}.
The second and third notes in the green voice overlap with the first one, and should have been put into a different voice.

\subsubsection{Measure Overflow}

In theory, the time signature constrains the duration of the measure, \ie it sets the \emph{nominal duration}. For example, a voice in a 4/4 measure is expected to contain the equivalent of four quarter notes.
In practice, as there are valid cases in which a measure is shorter than its nominal duration, \eg pickup measures or repetition bars placed in the middle of a measure.
Hence, we only treat measures longer than their nominal duration as errors.

\subsubsection{Chord Errors}

Chords are groups of notes belonging to the same voice, starting at the same time and lasting the same duration.
Once again, this is not constrained by the MusicXML specification. Therefore, we want to explicitly ensure that all same-voice notes starting at the same time also share the same duration, meaning that they belong to the same chord.

In addition, we can include some instrument-specific rules related to chords.
In the case of the piano dataset ASAP presented in \cref{sec:asap}, the piano is not allowed to play the same note multiple times simultaneously. Hence, chords in piano scores should not contain duplicate notes.\footnote{Yet, there can be duplicate notes if they belong to different voices.}

\section{Error Detection Methods}

We now introduce our error detection method, which is separated into two stages. The first stage checks for individual rhythm/time consistency errors as introduced in \cref{sec:individual-errors}, while the second one checks for contextual errors presented in \cref{sec:contextual-errors}.
Individual errors are straightforward to spot as they only require analyzing each element independently. However, contextual errors are more challenging, as they require considering the neighborhood of the tested elements.

Note that we focus on verifying the actual \emph{musical content} of the score.
In other words, we \textit{i)} consider only well-formed files, \ie respecting the MusicXML specification\footnote{\url{https://www.w3.org/2021/06/musicxml40}}, 
malformed files are supposed to be detected beforehand, 
and \textit{ii)} exclude engraving/rendering-related issues.

\subsection{Detection of Individual Errors}\label{sec:individual-errors-method}

We ensure the duration consistency of every note and rest, as explained in \cref{sec:individual-errors}.
First, we compute the theoretical duration of each of them from the \xmltag{type} and \xmltag{dot} tags,  taking into account any tuplet modifier tag \xmltag{time-modification}, if present. 
Second, we compare this value to the actual timeline duration of the element encoded in the MusicXML.
If they differ, we flag the note/rest as faulty for further inspection.

In addition to notes and rests, we also inspect \xmltag{backup} and \xmltag{forward} tags for suspicious durations.
To do so, we ensure that each of those tags can be decomposed into a valid sequence of rests, up to some user-defined minimal duration, like 128th notes.
For example, in a 480-division-based MusicXML file, a \xmltag{forward} of duration 45 could be decomposed into a 64th rest (30) and a 128th (15) rest, but a \xmltag{forward} of 46 could not without allowing much smaller divisions.
Any tag that cannot be decomposed into an allowed sequence of rests is flagged as faulty.

\subsection{Detection of Contextual Errors}\label{sec:contextual-errors-method}

\begin{table*}[t]
\rowcolors{1}{white}{gray!15}
\centering
\resizebox{\textwidth}{!}{%
\begin{tabular}{@{} ll p{0.9\textwidth} @{}}
\toprule
\textbf{Token type} &
  \textbf{Definition} &
  \textbf{Attributes} \\ \midrule
\token{Bar} &
  Beginning of a new measure &
  \begin{tabular}[c]{@{}l@{}}\tabitem $T_\text{numerator}$: Numerator of the time signature of the measure\\ \tabitem $T_\text{denominator}$: Denominator of the time signature of the measure\end{tabular} \\
\token{Note} &
  Single note &
  \begin{tabular}[c]{@{}l@{}}\tabitem $T_\text{voice}$: Voice index\\ \tabitem $T_\text{pitch}$: Pitch information (step, octave, accidental)\\ \tabitem $T_\text{duration}$: Symbolic duration (quarter note, half note, with optional dots\dots)\\ \tabitem $T_\text{is\_grace}$: Flag indicating if it is a grace note\\ \tabitem $T_\text{is\_tied}$: Flag indicating if tied to a future note\end{tabular} \\
\token{Rest} &
  Single rest &
  \begin{tabular}[c]{@{}l@{}}\tabitem $T_\text{voice}$: Voice index\\ \tabitem $T_\text{duration}$: Symbolic duration\\ \tabitem $T_\text{full\_measure}$: Flag indicating a full-measure rest\end{tabular} \\
\token{ChordStart} &
  Beginning of a chord &
  \begin{tabular}[c]{@{}l@{}}\tabitem $T_\text{voice}$: Voice index\\ \tabitem $T_\text{duration}$: Symbolic duration\\ \tabitem $T_\text{is\_grace}$: Flag indicating if it is a grace chord\end{tabular} \\
\token{ChordEnd} &
  End of a chord &
  \tabitem $T_\text{voice}$: Voice index \\
\token{TupletStart} &
  Beginning of a tuplet &
  \begin{tabular}[c]{@{}l@{}}\tabitem $T_\text{voice}$: Voice index\\ \tabitem $T_\text{base\_unit}$: Base symbolic unit\\ \tabitem $T_\text{n\_normal}$: Number of base units it should normally take (outside context)\\ \tabitem $T_\text{n\_actual}$: Number of base units it actually consists of (inside context)\\ \textit{\eg for a tuplet of 3:2 eighth notes, those last three attributes are respectively "eighth", 2 and 3.}\\ We also use the normal and actual duration of the tuplet:\\ \tabitem $T_\text{normal\_duration} = T_\text{base\_unit}\times T_\text{n\_normal}$\\ \tabitem $T_\text{actual\_duration} = T_\text{base\_unit}\times T_\text{n\_actual}$\end{tabular} \\
\token{TupletEnd} &
  End of a tuplet &
  \tabitem $T_\text{voice}$: Voice index \\
\token{EndOfScore} &
  End of the score &
  None \\ \bottomrule
\end{tabular}%
}
\caption{Token types and their attributes}
\label{tab:token-types}
\end{table*}

\begin{figure}
    \centering
    \includegraphics[width=\columnwidth]{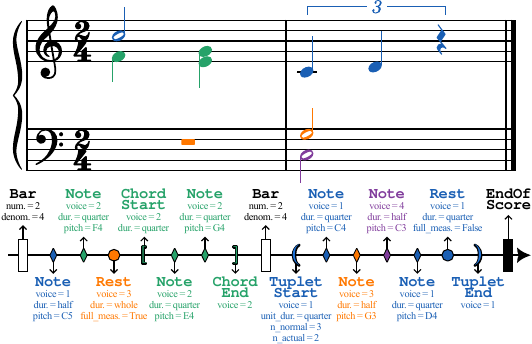}
    \vspace{-2em}
    \caption{Example of tokenization of a score. The timeline represents the sequence of tokens. The items' shape represents the type of token, and their color the voice they belong to (only for tokens tied to a voice).}
    \label{fig:tokenization}
\end{figure}

Detecting more complex errors than individual errors requires taking into account the interpretation context of each element.
We define a set of rules that scores must follow to be considered valid.
They are checked against a linearization of the score into a sequence of \textit{tokens}, representing different musical elements.
This sequence is parsed by a state machine that maintains a state about the partially analyzed token sequence, 
and checks the validity of each new token according to our ruleset.
We first detail how we tokenize the score, then the implementation of our state machine, and finally the rules that define allowed transitions.

\subsubsection{Tokenization of a Score}

We define several token types with attributes that represent various musical elements, \eg notes, rests, tuplets, bars, etc.
In this work, we focus on encoding elements relative to the rhythmic organization of the score.
However, note that this tokenization is generic and could be  extended to include other elements such as key signature and clef changes, \etc
All token types and their attributes are detailed in \cref{tab:token-types}.

The score is tokenized sequentially, going through musical symbols in a chronological order.
To avoid issues with the boundaries of repeated sections (with tied notes for example), 
we unfold all repetitions before tokenizing the score.
The order of simultaneous notes and rests does not have any impact on the ruleset we use.
Simultaneous notes in the same voice are wrapped in between \token{ChordStart} and \token{ChordEnd} tokens.
Similarly, in-tuplet notes and rests are wrapped in between \token{TupletStart} and \token{TupletEnd} tokens.
An example of such tokenization is given in \cref{fig:tokenization}.

\subsubsection{State Machine}

The next step consists in validating or invalidating a sequence of tokens according to a set of rules.
For this purpose, we use a state machine that tracks the state of the partially parsed sequence.
We call this state the \textit{score state}, notated $S$, containing two attributes:
\begin{itemize}
    \item $S_\text{voices}$ -- List of \textit{voice states}, describing the current completion of each voice.
    \item $S_\text{tied}$ -- Set containing all currently tied notes.
\end{itemize}
We note $S_0$ a special state representing the initial state of the machine, which will only be used later on to ensure 
that a score starts with a \token{Bar} token.

A voice state $v$ has the following attributes:
\begin{itemize}
    \item $v_\text{duration}$ -- Duration of the voice, in quarter notes, derived from the current time signature.
    \item $v_\text{position}$ -- Current position in the voice.
    \item $v_\text{chord}$ -- Either a \textit{chord state}, when inside a chord (\ie after a \token{ChordStart} token),
          or $\varnothing$ otherwise.
    \item $v_\text{tuplets}$ -- Stack of \textit{tuplet states}, representing potential tuplets in construction, 
    with the most nested one at the top of the stack. It is $\varnothing$ when there are no ongoing tuplets. 
    The function $\operatorname{top}$ retrieves the top element of the stack, \ie the most nested tuplet.
\end{itemize}

A chord state $C$ has the following attributes:
\begin{itemize}
    \item $C_\text{duration}$ -- Symbolic duration of the chord.
    \item $C_\text{notes}$  -- List of notes in the chord.
\end{itemize}

\begin{figure*}[h]
\centering
\begin{tabular}{cc}
     \adjustbox{valign=m}{
        \begin{tabular}{@{}c@{}}
            \subfloat[\token{Bar} token]{
                \begin{forest}
[\token{Bar}
    [OR
        [{$S = S_0$}]
        [AND, label={east:for $v \in S_\text{voices}$}
            [AND
                [{$v_\text{chord} = \varnothing$}]
                [{$v_\text{tuplets} = \varnothing$}]
                [{OR}
                    [{$v_\text{position} = 0$}]
                    [{$\displaystyle v_\text{position} = \max_{w \in S_\text{voices}}{w_\text{position}}$}]
                ]
            ]
            [...]
        ]
    ]
]
\end{forest}
                \label{fig:bar-condition}
            } \\
            \subfloat[\token{TupletStart} token]{
                \begin{forest}
[\token{TupletStart}
    [AND
        [{$S \neq S_0 $}]
        [{$v_\text{chord} = \varnothing$}]
        [OR  
            [AND
                [{$v_\text{tuplets} = \varnothing$}]
                [{$v_\text{position} + T_\text{normal\_duration} \leq v_\text{duration}$}]
            ]
            [AND
                [{$v_\text{tuplets} \neq \varnothing$}]
                [{$\topstack{v_\text{tuplets}}_\text{position} + T_\text{normal\_duration}$ \\ $\leq \topstack{v_\text{tuplets}}_\text{actual\_duration}$}]
            ]
        ]
    ]
]
\end{forest}
                \label{fig:tupletstart-condition}
            }
        \end{tabular}
    } & 
    \adjustbox{valign=m}{
        \subfloat[\token{Rest} token]{
            \begin{forest}
[\token{Rest}
    [AND
        [{$S \neq S_0 $}]
        [{$v_\text{chord} = \varnothing$}]
        [OR  
            [AND  
                [{$T_\text{is\_measure}$}]
                [AND  
                    [{$v_\text{position} = 0$}]
                    [{$v_\text{tuplets} = \varnothing$}]
                ]
            ]
            [AND  
                [{$\neg(T_\text{is\_measure})$}]
                [OR  
                    [AND  
                        [{$v_\text{tuplets} = \varnothing$}]
                        [{$v_\text{position} + T_\text{duration} \leq v_\text{duration}$}]
                    ]
                    [AND  
                        [{$v_\text{tuplets} \neq \varnothing$}]
                        [{
                            $\topstack{v_\text{tuplets}}_\text{position} + T_\text{duration}$ \\
                            $\leq \topstack{v_\text{tuplets}}_\text{actual\_duration}$
                        }]
                    ]
                ]
            ]
        ]
    ]
]
\end{forest}
            \label{fig:rest-condition}
        }
    }
\end{tabular}
\caption{Guard condition examples for a token $T$ for a state $S$, notating $v = S_\text{voices}[T_\text{voice}]$}
\label{fig:guard-conditions}
\end{figure*}
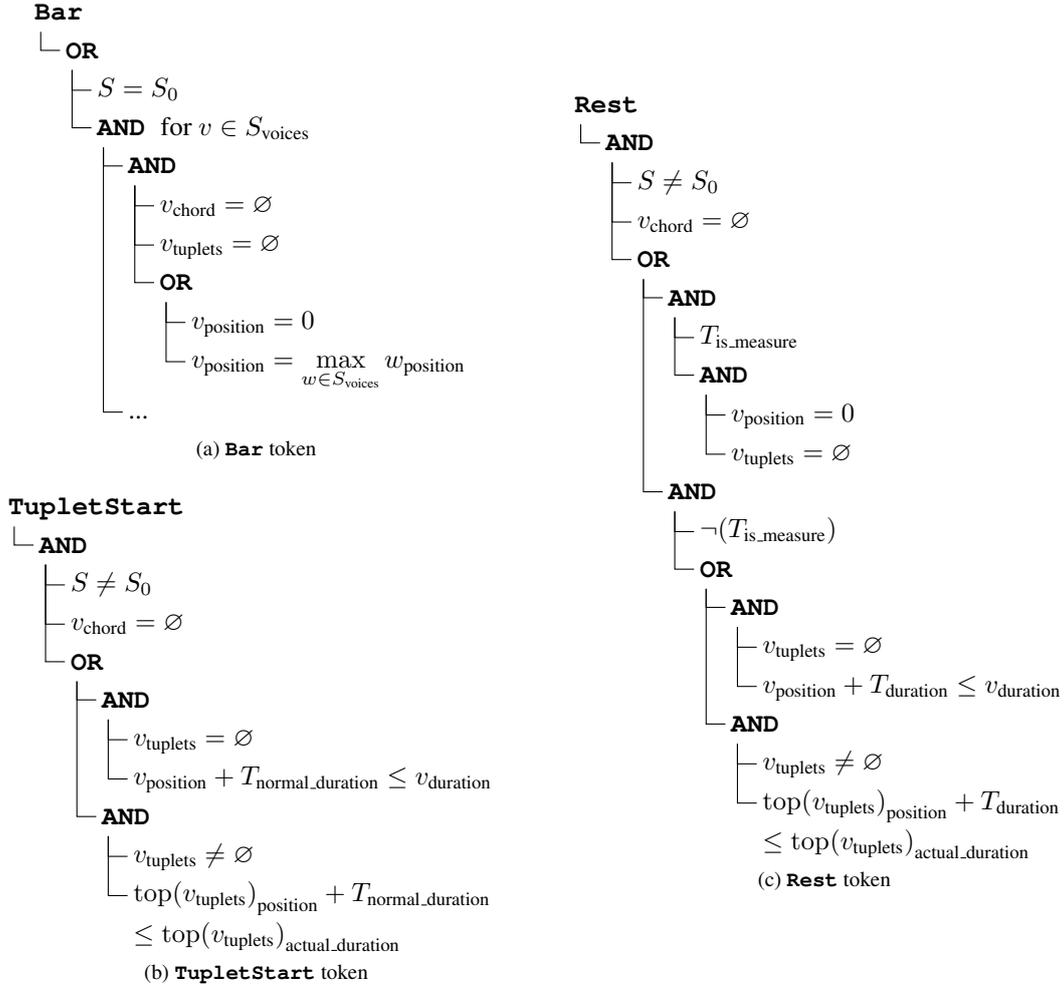

A tuplet state $K$ has the following attributes:
\begin{itemize}
    \item $K_\text{position}$ -- Current position in the tuplet (relative to the tuplet inner metric).
    \item $K_\text{unit\_duration}$ -- Base unit duration of the tuplet, \eg a eighth note in a 3:2 tuplet of eighth notes.
    \item $K_\text{n\_normal}$ -- Number of units that the tuplet should normally take. In the example above it would be 2.
    \item $K_\text{n\_actual}$ -- Number of units that the tuplet actually contains. In the example above it would be 3.
    \item $K_\text{normal\_duration} = K_\text{n\_normal} \times K_\text{unit\_duration}$
    \item $K_\text{actual\_duration} = K_\text{n\_actual} \times K_\text{unit\_duration}$
\end{itemize}

\subsubsection{Guards of Transitions}

Starting from the initial state $S_0$, we read the token sequence incrementally.
For each new token, we ensure that the transition is valid before updating the state,
by checking a verification condition called a \emph{guard}.
Each guard is token-specific and ensures that the current state can be updated by the new token through a valid transition.
We detail some guards in \cref{fig:guard-conditions}, while an exhaustive list is available in \cref{apx:transition-conditions}.
If a guard condition is false at some point when parsing a sequence, this means that the score does not respect one of our rules, which raises an error.

For example, a \token{Bar} token (\cref{fig:bar-condition}) is allowed either at the very beginning of the sequence ($S = S_0$), or if all voices are complete, \ie all chords have ended, all tuplets have ended, and all voices are synchronized\footnote{We recall that we do not consider incomplete voices as they have been filled with rests to fill the gaps.} or empty.

A \token{Rest} token (\cref{fig:rest-condition}) is \emph{not} allowed to come first in the sequence ($S \neq S_0$) and it cannot occur when a chord is in construction in the voice. 
To be valid, it must additionally be either a full-measure rest alone in the voice, 
or a regular rest which does not exceed the remaining duration of the voice -- or tuplet if there is an ongoing tuplet in the voice.

Similarly, a \token{TupletStart} token (\cref{fig:tupletstart-condition}) is \emph{not} allowed as the first token of the sequence and cannot occur if there is a chord in construction in the voice. 
In addition, the tuplet's normal duration must not exceed the remaining duration of the voice -- or outer tuplet if it is a nested tuplet.

Those guards could be extended to enforce stricter rules, or made optional to be more lenient with edge cases.

\subsubsection{State Update}

If the guard passes, then we can update the state of the machine. The current state is modified based on the new token and its attributes. These updates are generally straightforward.
For example, \token{Bar} resets all voice positions to 0 and sets all voice durations according to its time signature.
\token{Rest} advances the position of its voice by its duration.
\token{TupletStart} pushes a new tuplet state onto the tuplets stack of the voice, that will be popped out when a \token{TupletEnd} occurs on that same voice.

\section{Implementation and Application}\label{sec:application}

\begin{table*}[t]
\centering
\setlength{\tabcolsep}{4pt}
\begin{tabularx}{\textwidth}{rcccYYY}
\toprule
 &
  Scores w/ errors &
  \% scores w/ errors &
  Bars w/ errors &
  \% of bars w/ errors &
  Median bars w/ errors in flagged score\\
  \midrule
Individual errors &
  65 &
  $\frac{65}{235} = 28\%$ &
  692 &
  $\frac{692}{36496} = 1.9\%$ &
  3 \\
Contextual errors &
  35 &
  $\frac{35}{170} = 21\%$ &
  165 &
  $\frac{165}{24648^*} = 0.7\%$ &
  2 \\ \bottomrule
\end{tabularx}
\caption{Repartition of errors in ASAP dataset. The * indicates that the number of measures has been computed with unfolded scores, excluding the three incorrectly parsed MusicXML files mentioned in \cref{sec:check-of-contextual}.}
\label{tab:results}
\end{table*}

We assess the effectiveness of our error detection system by applying our syntactic checks on the MusicXML piano scores of the ASAP dataset \cite{foscarin:asap-dataset}.

\subsection{Implementation Details}

Our implementation is publicly available\footnote{\url{https://github.com/leleogere/detecting-notational-errors-in-digital-music-scores}}.

\subsubsection{Individual Errors}

For individual errors, we parse the MusicXML file and compare the different durations as explained in \cref{sec:individual-errors-method}.

\subsubsection{Tokenization}

We first parse the MusicXML files using Partitura \cite{cancino:partitura}, which keep the timeline untouched, meaning that overlapping notes in MusicXML still overlap in Partitura's internal representation.
We then iterate chronologically over all elements on the timeline, creating tokens for supported ones.

Using Partitura's representation as input for our tokenizer allows the contextual error detection method to support not only MusicXML, but also all other score formats supported by Partitura.
It could even be extended to unsupported formats by implementing the tokenization logic for those.

\subsubsection{State Machine}

We implement the states and guarded transitions introduced in \cref{sec:contextual-errors-method}.
While nested tuplets are allowed in our framework, we chose not to implement them as they should be extremely rare, and the ones in the ASAP dataset are mainly due to encoding errors anyway.
We also dropped support for grace chords, as Partitura currently does not support them. Instead, they are parsed as sequential grace notes. This has no practical impact on our method.

\subsection{Results on ASAP Dataset}\label{sec:asap}

ASAP \cite{foscarin:asap-dataset} contains 235 MusicXML scores, aligned with performances.
However, those scores were not written by professionals, and therefore include numerous mistakes and notational errors.
We report a summary of the proportion of the dataset impacted by both types of errors in \cref{tab:results}.

\subsubsection{Check of Individual Errors}

Among the 235 scores in ASAP, 65 are flagged for inconsistency between theoretical and actual duration of notes, representing around 28\% of the dataset.
Those errors are present in 692 measures, or 1.9\% of the total number of measures in the dataset.
Most erroneous scores contain only a few invalid measures, with the median number of bad measures per score being 3.
Only 17 scores have more than 5\% of measures containing this type of errors, and 5 more than 30\%.

\begin{figure}[t]
    \centering
    \includegraphics[width=1\linewidth]{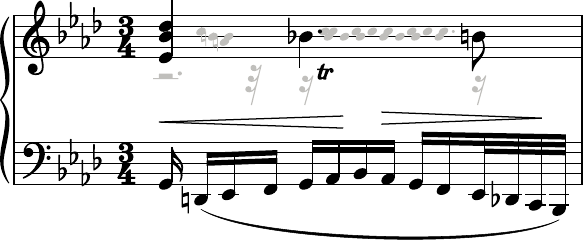}
    \vspace{-2em}
    \caption{Example of invisible notes (in gray) of arbitrary duration resulting in a non-decomposable \xmltag{backup} tag in measure 55 of the 1st movement of Beethoven's Sonata No. 31. Those notes were likely added for MIDI playback of the trill, and mess up with the MusicXML timeline.}
    \label{fig:individual-error-example-score}
\end{figure}

\Cref{fig:individual-error-example-xml,fig:individual-error-example-score} show some representative errors detected with this method.
Most of them are either rounding errors due to a bad division value in the MusicXML, or invisible notes with unusual durations included solely for MIDI playback.

\begin{figure}[t]
    \centering
    \inputminted[fontsize=\small]{xml}{figures/incorrect_scores/Beethoven_Piano_Sonatas_17-2_M94_individual-error.xml}
    \vspace{-1em}
    \caption{Example of tuplet rounding error in measure 94 of Beethoven's Sonata No. 17, op. 32, 2nd movement. The theoretical duration of a 32nd note in a 14:8 tuplet should be $\frac18 \times \frac8{14} \times 480 = \frac{240}7 \times 480 \approx 34.3 \neq 34$.}
    \label{fig:individual-error-example-xml}
\end{figure}

\subsubsection{Check of Contextual Errors}\label{sec:check-of-contextual}

Because our state machine only deals with theoretical durations, and not with MusicXML timeline durations, individual errors will trigger contextual errors too as voices will be inconsistently filled.
Therefore, we only run the contextual checks on the 170 correct scores left after the first stage, resulting in 35 incorrect scores:
\begin{itemize}
    \item One had two parts, which we did not implement as a piano score should only have one part (containing multiple staves, but still one part).
    \item Two were only partially parsed by Partitura, which had trouble detecting some tuplet durations, which our tokenization procedure relies on. Those parsing difficulties likely point towards a notational error.
    \item The 32 remaining scores had been flagged as invalid by our set of rules.
\end{itemize}

\begin{figure}[t]
    \centering
    \includegraphics[width=1\linewidth]{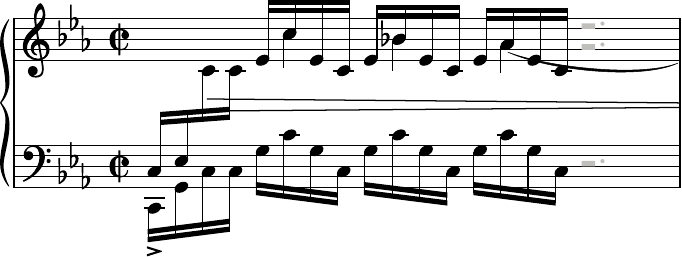}
    \vspace{-2em}
    \caption{Measure overflow due to invisible rests (in gray) in measure 7 of Chopin's Etude Op. 25 No. 12.}
    \label{fig:example-measure-overflow-rests}
\end{figure}

\begin{figure}[t]
    \centering
    \includegraphics[width=0.6\linewidth]{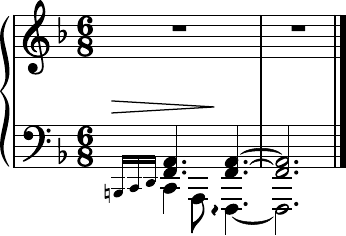}
    \vspace{-0.5em}
    \caption{Measure overflow due to cue notes that should be grace notes, bar 97 of Liszt's Transcendental Etude No. 3.}
    \label{fig:example-measure-overflow-grace}
\end{figure}

\begin{figure}[t]
    \centering
    \includegraphics[width=0.7\linewidth]{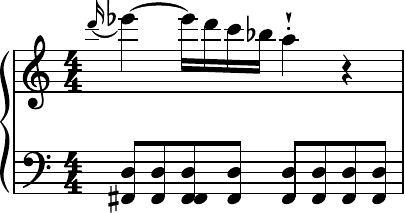}
    \vspace{-0.5em}
    \caption{Duplicated note in chord in measure 97 of the 1st movement of Beethoven's Piano Sonata No. 21.}
    \label{fig:example-duplicate-in-chord}
\end{figure}

Exploring those invalid scores, we successfully detected the following errors:
\begin{itemize}
    \item \textbf{Measure overflow:} a voice is too long for the current time signature. This is generally caused by invisible rests (\cref{fig:example-measure-overflow-rests}) or grace notes encoded as cue notes, which then have a non-zero duration (\cref{fig:example-measure-overflow-grace}).
    \item \textbf{Same-voice notes overlap:} a note starts before the end of a former note in the same voice. While not directly detected as an overlap, this will trigger a measure overflow as our state machine does not have a direct notion of position; instead, it adds notes in the voice, eventually resulting in an overflow. For instance, in \cref{fig:overlapping-notes}, the green voice is entirely filled by the first dotted quarter note, then the dotted eighth note coming next triggers an overflow as there is no more space in the measure.
    \item \textbf{Duplicate notes in chord:} as ASAP is a piano dataset, we forbid duplicated notes in a chord, as in \cref{fig:example-duplicate-in-chord}.
    \item \textbf{Different durations in chord:} no such error were detected in ASAP.\footnote{In practice, our implementation does not find any such error as they are already fixed by Partitura's MusicXML import. Nonetheless, we confirmed that no such error exists in ASAP by directly parsing the MusicXML files.}
\end{itemize}

\begin{figure}[t]
    \centering
    \includegraphics[width=\linewidth]{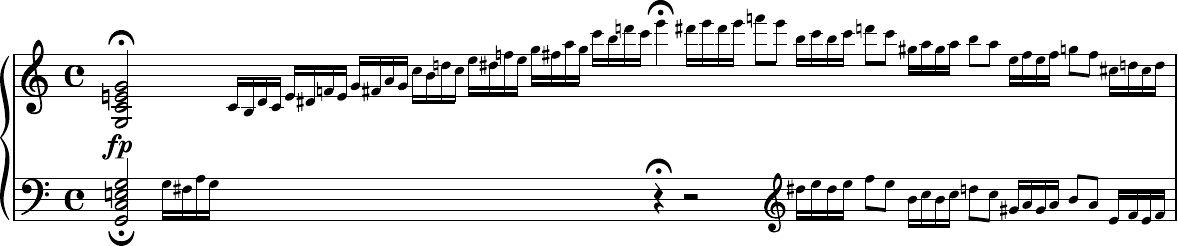}
    \vspace{-1.5em}
    \caption{Free-played section in measure 232 of the 1st movement of Beethoven's Sonata No. 3. The actual duration of the measure is far longer than its nominal duration.}
    \label{fig:example-free-play}
\end{figure}

Note that the measure-overflowing mechanism also detects some occurrences of free played sections, such as in measures 232 to 234 of 1st movement of Beethoven's Piano Sonata No. 3 (\cref{fig:example-free-play}).
Those are considered as invalid by our model, as they go far beyond the nominal duration of the measure.
These instances could be considered acceptable, and therefore, flagging them constitutes a false positive.
This could be solved by allowing voices that are longer than the nominal measure duration, \ie, making the overflowing rule optional.
However, we argue this case (and other similar ones) is debatable: these small notes are encoded as cue notes, which are supposed to be unplayed notes solely present to inform the performer about some other instrument's melody, not a free-play indication.

\subsubsection{Fixing Scores}

After detecting scores with notational errors, we have manually fixed some and pushed them to the ASAP repository.\footnote{\url{https://github.com/fosfrancesco/asap-dataset}, only available on the \texttt{develop} branch as of October 2025.}
Two of them contained tuplet rounding errors, which were fixed by computing an adequate value for the \xmltag{divisions} tag, permitting an exact representation of all durations present in the score. All \xmltag{duration} tags were updated according to this new value.
Two scores were missing tuplet start tags, which would cause measure overflow issues later on.
They were fixed by adding them back.
Two scores were using incorrect tuplets, resulting in time-shifting and measure overflowing.
One score contained extra invisible rests in an already-full measure, so we removed them.
Finally, four other scores have also been fixed for other unrelated issues (extra staves, notes wrongly marked as invisible, poor notation choices).

\section{Discussion}

In this article, we introduce a two-step method for the detection of notation error in digital music scores.
The first step, specifically tailored for MusicXML, consists of ensuring consistency between different redundant duration information in MusicXML elements, in order to detect individual note errors.
The second step, format-agnostic, relies on an implemented tokenizer which linearizes the score into a sequence of tokens. This sequence is then checked by a state machine, whose role is to test the validity of each token within the musical surrounding context, in order to detect contextual errors.

Combining those two steps, our method is able to detect errors in 100 of the 235 scores of the ASAP dataset, or around 42\%, confirming the fact that these scores were not written by professionals, and highlighting the interest of such detection methods.
This allowed us to fix multiple scores in the dataset, enhancing its global quality.
Scores that were manually fixed have been backported to the ASAP repository, for others to benefit from our work, even though there is still a long way to go to get a clean dataset.

Our method and implementation are modular and could be easily extended with more rules encoding various score quality aspects \cite{foscarin:data-quality2021}, to identify other types of errors.
For instance, we might want to ensure that a score respects an specific instrument's tessitura, or make sure that no tie crosses a key signature or clef change -- while not strictly forbidden, it is confusing to read.
Hence, this model might serve as a base to build a more complete score validator to assess the quality of both human-produced and computer-generated music scores.

The main limitation of our approach is that while we detect errors, we do not automatically fix them.
It is possible to suggest fixes for some error types, such as proposing a more suitable \xmltag{duration} tag to correct duration inconsistencies. However, this still requires manual intervention and some expertise to check that the fix is suitable.
Indeed, to fix rhythm/time consistency errors on notes and rests, one would need to choose which information becomes authoritative: the theoretical duration computed from the symbol or the timeline duration.
The \xmltag{forward}/\xmltag{backup} tags, introduced to keep up with alignment in case of rounding errors, are even more difficult to deal with.
Conversions between software can introduce invisible rests that are difficult to clean up automatically.
Contextual errors are especially hard to automatically fix, because they require understanding the intention of the composer to sort out the cause of the error (\eg, why is a measure overflowing?) and some expertise to know how to correct it while following the conventions of music notation.
We observed that mistakes often tend to be very specific to one score, as two individuals have different ways of transcribing a piece.
While a fully automated correction pipeline might be out of reach, our work remains useful to flag issues in scores.
It then remains up to the user to inspect the problematic measures and determine the most appropriate fixes, \eg, by comparing to references of that score in other published editions if they exist.

\bibliography{references}

\pagebreak
\appendix
\onecolumn

\section{Example of a MusicXML containing Notational Errors}\label{apx:error-musicxml}

The XML in \cref{fig:incorrect-musicxml-source} is the source of the examples from \cref{fig:incorrect-musicxml}.
There are some duration inconsistencies in notes, as well as unusual \xmltag{backup}/\xmltag{forward} tags crossing the measure boundaries.
Different pieces of software deal with those differently, resulting in rendering differences.
This illustrates how the MusicXML specification does not enforce strict respect of musical notation, and allows for ambiguous interpretation when engraving.

\begin{figure}[h]
\inputminted[fontsize=\footnotesize,baselinestretch=0.95,breaklines]{xml}{figures/broken_musicxml_example/incorrect_score_source_reduced.xml}
\label{fig:incorrect-musicxml-source}
\caption{XML source of \cref{fig:incorrect-musicxml}.}
\end{figure}

\newpage

\section{Guard Conditions}\label{apx:transition-conditions}

We detail below the guard conditions for the token types that are not introduced in \cref{sec:contextual-errors-method}.
In each figure, we test whether the newly read token $T$ allows for a valid transition while the state machine is in the state $S$.

\begin{multicols}{2}

\begin{figure}[H]
\begin{forest}
[\token{Note}
    [AND
        [{$S \neq S_0 $}]
        [OR  
            [AND
                [{$v_\text{chord} \neq \varnothing$}]
                [{$T_\text{duration} = \left(v_\text{chord}\right)_\text{duration}$}]
                [{$T_\text{is\_grace} = \left(v_\text{chord}\right)_\text{is\_grace}$}]
            ]
            [AND
                [{$v_\text{chord} = \varnothing$}]
                [OR  
                    [AND
                        [{$T_\text{is\_grace}$}]
                        [{$v_\text{position} < v_\text{duration}$}]
                    ]
                    [AND
                        [{$\neg\left(T_\text{is\_grace}\right)$}]
                        [OR  
                            [AND
                                [{$v_\text{tuplets} = \varnothing$}]
                                [{$v_\text{position} + T_\text{duration} \leq v_\text{duration}$}]
                            ]
                            [AND
                                [{$v_\text{tuplets} \neq \varnothing$}]
                                [{
                                    $\topstack{v_\text{tuplets}}_\text{position} + T_\text{duration} $ \\
                                    $\leq \topstack{v_\text{tuplets}}_\text{actual\_duration}$
                                }]
                            ]
                        ]
                    ]
                ]
            ]
        ]
    ]
]
\end{forest}
\caption{Guard condition for a \token{Note} token}
\end{figure}
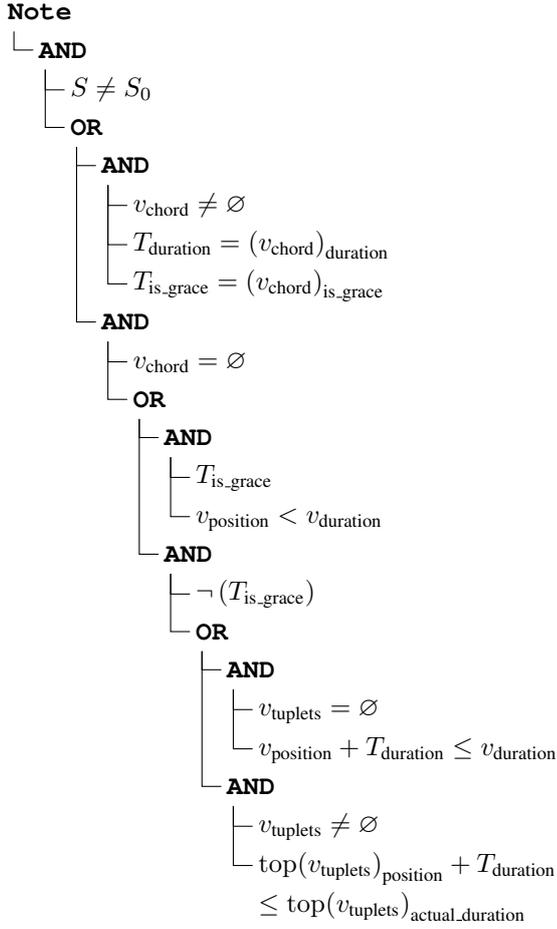

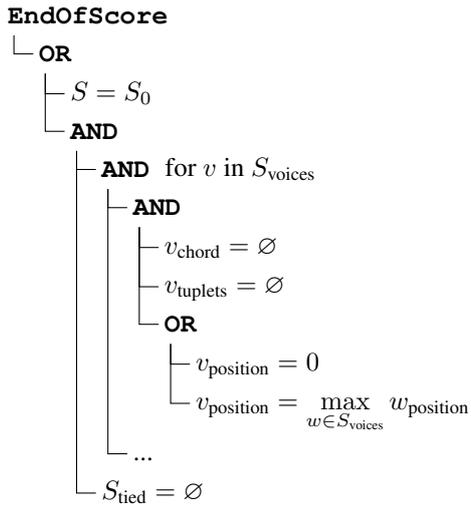
\begin{figure}[H]
\begin{forest}
[\token{EndOfScore}
    [OR
        [{$S = S_0 $}]
        [AND
            [AND, label={east:for $v$ in $S_\text{voices}$}
                [AND
                    [{$v_\text{chord} = \varnothing$}]
                    [{$v_\text{tuplets} = \varnothing$}]
                    [{OR}
                        [{$v_\text{position} = 0$}]
                        [{$\displaystyle v_\text{position} = \max_{w \in S_\text{voices}}{w_\text{position}}$}]
                    ]
                ]
                [...]
            ]
            [{$S_\text{tied} = \varnothing$}]
        ]
    ]
]
\end{forest}
\caption{Guard condition for an \token{EndOfScore} token}
\end{figure}

\begin{figure}[H]
\begin{forest}
[\token{ChordStart}
    [AND
        [{$S \neq S_0 $}]
        [{$v_\text{chord} = \varnothing$}]
        [OR  
            [AND
                [{$T_\text{is\_grace}$}]
                [{$v_\text{position} < v_\text{duration}$}]
            ]
            [AND
                [{$\neg\left(T_\text{is\_grace}\right)$}]
                [OR  
                    [AND
                        [{$v_\text{tuplets} = \varnothing$}]
                        [{$v_\text{position} + T_\text{duration} \leq v_\text{duration}$}]
                    ]
                    [AND
                        [{$v_\text{tuplets} \neq \varnothing$}]
                        [{
                            $\topstack{v_\text{tuplets}}_\text{position} + T_\text{duration}$ \\
                            $\leq \topstack{v_\text{tuplets}}_\text{actual\_duration}$
                        }]
                    ]
                ]
            ]
        ]
    ]
]
\end{forest}
\caption{Guard condition for a \token{ChordStart} token}
\end{figure}
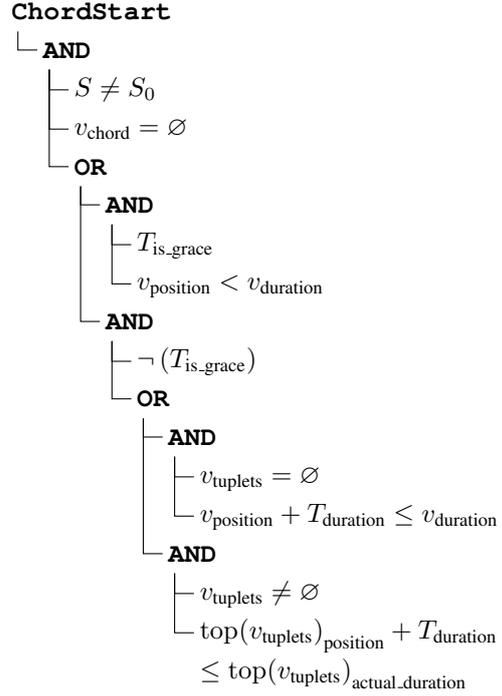

\begin{figure}[H]
\begin{forest}
[\token{ChordEnd}
    [AND
        [{$v_\text{chord} \neq \varnothing$}]
        [{$\left| \left(v_\text{chord}\right)_\text{num\_notes} \right| \geq 2$}]
    ]
]
\end{forest}
\caption{Guard condition for a \token{ChordEnd} token}
\end{figure}
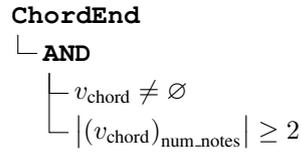

\begin{figure}[H]
\begin{forest}
[\token{TupletEnd}
    [AND
        [{$v_\text{tuplet} \neq \varnothing$}]
        [{$\topstack{v_\text{tuplets}}_\text{position} = \topstack{v_\text{tuplets}}_\text{duration}$}]
    ]
]
\end{forest}
\caption{Guard condition for a \token{TupletEnd} token}
\end{figure}
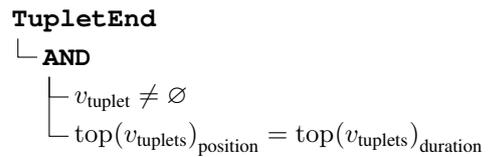
    
\end{multicols}

\end{document}